\documentclass[twocolumn,preprintnumbers,amsmath,amssymb,floatfix]{revtex4}
\input{epsf}

\usepackage{natbib}
\usepackage[dvips]{graphicx}
\usepackage{mciteplus}

\usepackage{subfigure}
\usepackage{dcolumn}
\usepackage{bm}
\usepackage{latexsym}

\newcommand{\ie}{\textsl{i.e.~}}

\def\spose#1{\hbox to 0pt{#1\hss}}
\def\lta{\mathrel{\spose{\lower 3pt\hbox{$\mathchar"218$}}
     \raise 2.0pt\hbox{$\mathchar"13C$}}}
\def\gta{\mathrel{\spose{\lower 3pt\hbox{$\mathchar"218$}}
     \raise 2.0pt\hbox{$\mathchar"13E$}}}

\newcommand{\mpl}{m_\mathrm{Pl}}
\newcommand{\de}[2]{\kern - #1 em \mathrm{d} #2}

\begin{document}

\title{Generation of gravitational waves during early structure
  formation between cosmic inflation and reheating}

\author{Karsten Jedamzik} \email{jedamzik@lpta.univ-montp2.fr}
\affiliation{Laboratoire de Physique Th\'eorique et Astroparticules, 
UMR 5207-CNRS, Universit\'e de Montpellier II, F-34095 Montpellier, France}

\author{Martin Lemoine} \email{lemoine@iap.fr} \affiliation{Institut
d'Astrophysique de Paris, \\ UMR 7095-CNRS, Universit\'e Pierre et Marie
Curie, \\ 98bis boulevard Arago, 75014 Paris, France}

\author{J\'er\^ome Martin} \email{jmartin@iap.fr}
\affiliation{Institut d'Astrophysique de Paris, \\ UMR 7095-CNRS,
Universit\'e Pierre et Marie Curie, \\ 98bis boulevard Arago, 75014
Paris, France}

\date{\today}

\begin{abstract}
  In the pre-reheating era, following cosmic inflation and preceding
  radiation domination, the energy density may be dominated by an
  oscillating massive scalar condensate, such as is the case for $V
  = m^2\phi^2/2$ chaotic inflation. We have found in a previous paper
  that during this period, a wide range of sub-Hubble scale
  perturbations are subject to a preheating instability, leading to
  the growth of density perturbations ultimately collapsing to form
  non-linear structures. We compute here the gravitational wave signal
  due to these structures in the linear limit and present estimates
  for emission in the non-linear limit due to various effects: the
  collapse of halos, the tidal interactions, the evaporation during
  the conversion of the inflaton condensate into radiation and finally
  the ensuing turbulent cascades. The gravitational wave signal could
  be rather large and potentially testable by future detectors.
\end{abstract}

\maketitle

\section{Introduction}
\label{sec:introduction}

One of the most salient features of inflationary cosmology is
undoubtedly the generation of both scalar and tensor modes of metric
perturbations on super-horizon scales through the amplification of
vacuum fluctuations. Scalar modes correspond to genuine density
fluctuations which are probed with ever increasing accuracy by the
measurements of the temperature fluctuations of the cosmic microwave
background and large scale surveys. Tensor modes, however remain to be
detected. Current (e.g. Planck~\cite{Bouchet:2009tr}) or next
generation experiments (e.g. BBO~\cite{Corbin:2005ny},
DECIGO~\cite{Kawamura:2006up}) should in principle detect the relic
gravitational wave background if the energy scale of inflation is
close to the GUT scale. As is well known, gravitational waves are much
harder to detect because of the intrinsic weakness of gravity, but at
the same time and for the same reason they offer an invaluable probe
of early Universe physics.

\par

In the past two decades, it has been realized that gravitational waves
could also be produced in the primordial Universe and on sub-horizon
scales due to a rapid disturbance of an otherwise equilibrium
homogeneous state. This may occur for instance in first order
primordial phase
transitions~\cite{Kosowsky:1992rz,Kosowsky:1991ua,Kosowsky:1992vn,Kamionkowski:1993fg,Nicolis:2003tg,Grojean:2006bp},
through bubble collisions and the dissipation of energy in turbulent
cascades~\cite{Kosowsky:2001xp,Dolgov:2002ra,Gogoberidze:2007an}, or
via the interactions of waves generated by parametric amplification in
preheating scenarios involving multiple scalar fields or tachyonic
couplings of the
inflaton~\cite{Khlebnikov:1997di,Easther:2006gt,Easther:2006vd,GarciaBellido:2007dg,GarciaBellido:2007af,Dufaux:2007pt}. These
various scenarios lead to different predictions at different
frequencies depending on the detailed processes at work, with
interesting prospects for detection~\cite{Maggiore:1999vm}.

\par

In a previous paper (hereafter Paper~I)~\cite{Jedamzik:2010dq}, we have shown
that {\em single field} inflation generically leads to the growth of
density fluctuations in the pre-reheating era for a certain range of
wave numbers on sub-Hubble scales, and we have argued that one of the
possible signatures of this effect would be the generation of
gravitational waves (GW). The objective of the present paper is to
calculate explicitly the spectrum of gravitational waves emitted by
the interactions of these density perturbations. In order to avoid
unnecessary model dependencies, we focus on the simplest model of
inflation, namely chaotic inflation with potential
$V(\phi)=m^2\phi^2/2$. Then, the normalization of the density
perturbation power spectrum to the value measured by cosmic microwave
probes sets the scale for the mass $m$ and the Hubble parameter at the
end of inflation. The only parameter left in the model is the
reheating temperature of the Universe, $T_{\rm rh}$, which marks the
beginning of the radiation dominated era and which is directly related
to the decay constant $\Gamma_\phi$ of the inflaton, $T_{\rm rh}\sim
0.2\sqrt{\Gamma_\phi M_{\rm Pl}}$ (with $M_{\rm Pl}$ the reduced
Planck mass). In the following, we calculate the present day
gravitational wave spectrum in the linear regime and provide detailed
estimates for gravitational wave emission due to non-linear effects,
which are associated with the growth of the density contrast on
different scales beyond unity. We find potentially observable signals,
all the more so when the reheating temperature is lowered. A lower
reheating temperature is associated with a longer epoch during which
density fluctuations may grow as well as a frequency closer to the
range of maximum sensitivity of next generation gravitational wave
detectors.

\par

Recently, the problem of gravitational wave emission by the growth of
density fluctuations in an early matter era has been discussed by
Assadullahi and Wands~\cite{Assadullahi:2009nf}. The problem at hand
is very similar but there are several differences between this study
and the present one. Firstly, Assadullahi and Wands do not specify the
mechanism through which fluctuations can grow but assume that all
sub-horizon modes are unstable if the equation of state of the
background field is on average pressure-less, as happens for
$m^2\phi^2$ during the inflaton oscillatory phase. As demonstrated in
Paper~I however, growth of fluctuations is a preheating effect, which
is limited to a band of wave numbers, whose range depends on the
precise shape of the potential. Borrowing from this calculation, we
are thus able in the present paper to relate the growth of
fluctuations with the initial post-inflationary power spectrum of
metric fluctuations. More importantly, we provide in the present paper
detailed estimates for the non-linear regime of matter fluctuations,
while the authors of Ref.~\cite{Assadullahi:2009nf} have focused their
study only on the linear regime.

\par

The outline of the paper is as follows: In
Section~\ref{sec:gravinstability} we recall the main findings of
Paper~I and set the stage for the calculation of the gravitational
wave signal done in Section~\ref{sec:grav-lin} in the linear limit.
Section~\ref{sec:nonlin} discusses in four subsections, gravitational
wave emission in the non-linear regime, due to the first collapse
phase of density fluctuations, due to the non-linear phases after
first collapse, due to the evaporation of the collapsed halos at
reheating and due to the dissipation of energy in turbulent cascades
after reheating, respectively. Conclusions are drawn in
Section~\ref{sec:conclusion}.

\section{Gravitational instability in the pre-reheating era}
\label{sec:gravinstability}

In this section, we briefly recall the main findings of Paper~I. In
that article, we have discussed the evolution of the Mukhanov variable
$v_\mathbf{k}$ and of the curvature perturbation $\zeta _\mathbf{k}$
during the pre-heating epoch. These two variables control the dynamics
of the scalar metric perturbations. During pre-heating, the inflaton
field oscillates according to
\begin{equation}
\phi (t)\simeq \phi_{\rm end}\left(\frac{a_{\rm end}}{a}
\right)^{3/2}\sin\left(mt\right)\, ,
\end{equation}
where $\phi_{\rm end}\simeq \mpl/(2\sqrt{3})$ is the inflaton vacuum
expectation value at the end of inflation, and where 
$\mpl \equiv \sqrt{8\pi} M_{\rm Pl}$. The mass $m$ is fixed by
the COBE/WMAP normalization, $m\simeq 1.7\times 10^{13}\,$GeV. The
Mukhanov variable $v_{\bm{k}}$ is related to the curvature
perturbation $\zeta_{\bm{k}}$ by
\begin{equation}
\zeta_{\bm{k}}=\sqrt{\frac{\kappa}{2}}\frac{v_{\bm{k}}}{a\sqrt{\epsilon
  _1}}\ ,
\end{equation}
where $\epsilon _1\equiv -\dot{H}/H^2$ is the first slow-roll
parameter expressed in terms of the Hubble parameter
$H=\dot{a}/a$. The quantity $a$ denotes the
Friedman-Lema\^{\i}tre-Robertson-Walker scale factor, $\kappa \equiv
M_{\rm Pl}^{-2}$ and a dot means a derivative with respect to cosmic
time.

\par

In Paper~I, we have shown that there exists an instability regime of
narrow resonance provided the wave number of the Fourier mode is such
that
\begin{equation}
0<\frac{k}{a}<\sqrt{3Hm}\, .
\end{equation}
Once a mode has entered this resonance band, it remains inside it
during the whole oscillatory phase. Moreover, we have also
demonstrated that, in the resonance band, $v_{\bm{k}}\propto a$
and, therefore, given the definition of this quantity given above,
that $\zeta_{\bm{k}}$ remains constant. Note that usually the
curvature perturbation remains constant only on super Hubble scales
while decaying on sub Hubble scales, unless there is gravitational
instability.  By using the perturbed Einstein equation
\begin{equation}
\label{eq:deltak}
\delta_{\bm{k}}=-\frac{2}{5}
\left(\frac{k^2}{a^2H^2}+3\right)\zeta _{\bm{k}}\, ,
\end{equation}
with the shorthand notation $\delta_{\bm{k}}(\eta)\,\equiv\,\delta
\rho_{\bm{k}}/\rho$ introducing the fractional mass-energy density
perturbation, and given that $\rho \propto 1/a^3$, we may see that
$\delta_{\bm{k}}$ grows as $a(t)$ on sub-Hubble scales and remains
constant on super-Hubble scales. The preheating instability for all
sub-Hubble modes with $k/a <\sqrt{3Hm}$ may be understood as the
gravitational instability of the oscillating scalar condensate.

\par

As already stated, the amplification described above occurs only for a
limited range of scales, $k\in \left[k_{\rm min},\,k_{\rm max}\right]$
in terms of comoving wave number. We now discuss this point in more
detail and specify $k_{\rm min}$ and $k_{\rm max}$. A priori, the
smallest unstable scale (in wavelength) is that which enters the
resonance band at the end of reheating. However, it has never been
outside the Hubble scale during inflation and, therefore, the power
stored on that scale at the end of inflation is suppressed. As a
consequence, for practical purposes, one can consider that the
smallest scale for which the amplification occurs is such that $k
\,\simeq\,a_{\rm end}H_{\rm end}$, \ie $\hat k \equiv k/(a_{\rm
  end}H_{\rm end})\simeq 1$, describing modes that leave and re-enter
immediately the Hubble radius at the end of inflation, hence $k_{\rm
  max}=a_{\rm end}H_{\rm end}$ ($\hat k_{\rm max}=1$). More detail on
this issue can also be found in paper~I.

\par

On the other hand, the largest scale is the one for which modes
re-enter the Hubble radius at the end of the matter dominated
pre-reheating epoch, \ie $k_{\rm min}\simeq a_{\rm rh}H_{\rm rh}$,
where $H_{\rm rh}$ is the Hubble parameter at the end of the reheating
epoch or at the beginning of the radiation dominated phase. One can
show that the number of e-folds between the Hubble radius exit of
$k_{\rm min}$ during inflation and the end of inflation is formally given by:
\begin{equation}
{\rm ln}\left(\frac{a_{\rm end}}{a_{k_{\rm min}}}\right)
=-\frac{1}{2}-\frac{1}{2}
W_{-1}\left(-e^{-1}\hat k_{\rm min}^2\right)\, ,
\end{equation}
where $W_{-1}$ denotes the Lambert function. If one neglects the
evolution of the Hubble scale over the interval between Hubble radius exit of
$k_{\rm min}$ and the end of inflation, the above can be approximated
with $a_{k_{\rm min}}/a_{\rm end}\simeq 
  \hat k_{\rm min}$.  In terms of the reheating
temperature and inflationary scale given by
\begin{eqnarray}
\hat k_{\rm min}&\, = \,& \frac{a_{\rm
      rh}H_{\rm rh}}{a_{\rm end}H_{\rm end}}\nonumber\\
&\,\simeq\,&  5.9\times10^{-5}\left(\frac{T_{\rm rh}}{10^9\,{\rm
      GeV}}\right)^{2/3}\ .\label{eq:kmin}
\end{eqnarray}
The last equation assumes $g_*=230$ degrees of freedom at reheating.
For a reheating temperature of $T_{\rm rh}\simeq 10^9\,$GeV, one finds
${\rm ln}(a_{\rm end}/a_{k_{\rm min}})\simeq 11$. Therefore, 
for $T_{\rm rh}\simeq 10^9\,$GeV, the
scales we would be interested in in this article are those which left the
Hubble radius during inflation between eleven and zero e-folds before
the end of inflation.

\par

\begin{figure}
  \centering
  \includegraphics[width=0.5\textwidth,clip=true]{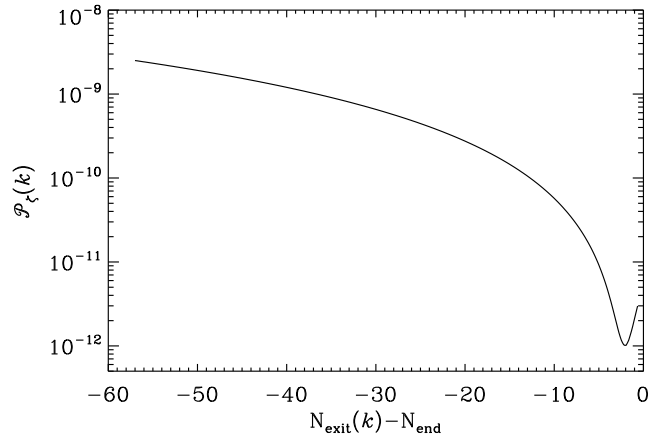}
  \caption[...]{Primordial power spectrum of curvature perturbations
    for the large field inflation model with
    $V(\phi)=m^2\phi^2/2$. Here $N_{\rm
      exit}(k)-N_{\rm end}$ is ${\rm ln}\,[a_{\rm cross}(k)/a_{\rm
      end}]$, with $a_{\rm cross}(k)$ the scale factor at Hubble
    radius crossing of mode $k$ during inflation and $a_{\rm end}$ the
    scale factor at the end of inflation.  } \label{fig:primspectre}
\end{figure}

Finally, we also need the initial power stored on each scale $k\in
\left[k_{\rm min},\,k_{\rm max}\right]$ at the end of inflation. This
is described by the power spectrum of the curvature fluctuation
defined by
\begin{equation}
{\cal P}_{\zeta}(k)\equiv \frac{k^3}{2\pi ^2}
\left\vert \zeta_{\bm{k}}\right\vert ^2
=\frac{2k^3}{\pi \mpl^2}\left\vert 
\frac{v_{\bm{k}}}{a\sqrt{\epsilon _1}}\right\vert ^2\, .
\label{Pzeta}
\end{equation}
The corresponding primordial power spectrum for $m^2\phi^2/2$ inflation
correctly normalized to COBE/WMAP observations is shown
in Fig.~\ref{fig:primspectre}. This figure has been obtained through a
numerical integration of the evolution of $\zeta_{\bm{k}}$ (through
$v_{\bm{k}}$): as discussed in Paper~I, the slow-roll approximation
at first or even second order cannot reliably predict the shape of the
power spectrum in the range of comoving wave numbers of interest to us,
$k\,\lesssim\, a_{\rm end}H_{\rm end}$, which exit and re-enter the
Hubble radius not too long afterward (in particular, before reheating
is completed). In the notation of Fig.~\ref{fig:primspectre}, $N_{\rm
  exit} -N_{\rm end}\sim -10\rightarrow 0$, the lower bound depending
on the exact value of the reheating temperature.

\section{Generation of Gravitational Waves after Inflation: 
the Linear Regime}
\label{sec:grav-lin}

Let us now consider the production of GWs induced by the phenomenon
described in the previous section. Our analysis follows the main
thread of the analysis presented in
Refs.~\cite{Kosowsky:2001xp,Dufaux:2007pt}. We work with the metric
\begin{equation}
{\rm d}s^2 = a^2\bigl\{(1+2\Phi ){\rm d}\eta^2 
- \bigl[(1+2\Phi )\delta_{ij}+h_{ij}\bigr]{\rm d}x_i{\rm d}x_j\bigr\}\, ,
\label{metric}
\end{equation}
where $\Phi$ is the gauge-invariant potential (\ie the potential in
longitudinal gauge) which describes the presence of inhomogeneities in
the inflaton field and $h_{ij}$ is the transverse traceless spatial
tensor accounting for the generation of GWs.  We are interested in the
amplitude $h_{ij}$ of the transverse and traceless which obey the
following equation (in Fourier space)
\begin{equation}
\label{eq:eomgw}
\bar{h}_{ij}''+\left(k^2-\frac{a''}{a}\right)\bar{h}_{ij}
=2\kappa a T_{ij}^{_{\rm TT}}(\eta ,\bm{k})\, ,
\end{equation}
where we used the rescaled amplitude $\bar{h}_{ij}=ah_{ij}$ and with
$T_{ij}(\eta,\bm{k})=\int {\rm d}\bm{x}\, {\rm
  e}^{i\bm{k}\cdot \bm{x}}T_{ij}(\eta
,\bm{x})/(2\pi)^{3/2}$. The transverse traceless part
of the spatial energy-momentum tensor $T_{ij}^{_{\rm TT}}$ is obtained by
\begin{equation}
T_{ij}^{_{\rm TT}}\equiv \perp_{ij\ell m}T_{\ell m}\, ,
\end{equation}
with
\begin{eqnarray}
\label{eq:defperp}
\perp_{ij\ell m}\equiv P_{i\ell}(\bm{e}_{\bm{k}})
P_{jm}(\bm{e}_{\bm{k}})
-\frac{1}{2}P_{ij}(\bm{e}_{\bm{k}})
P_{\ell m}(\bm{e}_{\bm{k}})\, ,
\end{eqnarray}
and
\begin{eqnarray}
\label{eq:defP}
P_{ij}\left(\bm{e}_{\bm{k}}\right)\equiv \delta _{ij}-e_{\bm{k},i}
e_{\bm{k},j}\, .
\end{eqnarray}
Here the quantity $\bm{e}_{\bm{k}}\equiv \bm{k}/k$ is the unit vector
in the $\bm{k}$ direction. We are interested in the generation of GWs
by the parametrically growing inflaton density perturbations between
the end of inflation and the epoch of reheating, \ie the epoch of
inflaton decay.  We will see below that the lowest non-vanishing
contribution of these perturbations to $T_{ij}^{_{\rm TT}}$ appears
only at second order in the perturbative quantity
$\delta_{\bm{k}}$. We will thus work to this order in $T_{ij}^{_{\rm
    TT}}$ but only to first order in all other quantities.  If we
neglect the term $a''/a$, as appropriate for sub Hubble scales, then
the solution of the above equation can be written as
\begin{equation}
\bar{h}_{ij}(\eta ,\bm{k})=A_{ij}\left(\bm{k}\right)
\sin \left[k\left(\eta -\eta _{\rm f}\right)\right]
+B_{ij}\left(\bm{k}\right)
\cos \left[k\left(\eta -\eta _{\rm f}\right)\right]\, ,
\end{equation}
where
\begin{eqnarray}
\label{eq:A}
A_{ij} &=& \frac{2\kappa }{k}
\int _{\eta _{\rm ini}}^{\eta_{\rm f}}{\rm d}\tau
\cos \left[k\left(\eta _{\rm f}-\tau \right)\right]a(\tau)
T_{ij}^{_{\rm TT}}\left(\tau,\bm{k}\right)\, ,\\
\label{eq:B}
B_{ij} &=& \frac{2\kappa }{k}
\int _{\eta _{\rm ini}}^{\eta_{\rm f}}{\rm d}\tau
\sin \left[k\left(\eta _{\rm f}-\tau\right)\right]a(\tau)
T_{ij}^{_{\rm TT}}\left(\tau,\bm{k}\right)\, . 
\end{eqnarray}
In these expression, we have assumed that the source is non negligible
in the time interval $\eta _{\rm ini}<\eta <\eta _{\rm f}$. In our
calculation, the initial time is set by the time at which a given mode
enters the Hubble radius, while $\eta_{\rm f}$ is given by the time at
which a mode reaches non-linearity. A discussion of generation of GWs
in the non-linear regime is deferred to the next section.  The energy
density of GWs is given by
\begin{widetext}
\begin{equation}
\rho _{\rm gw}(\eta )
=\frac{1}{4\kappa a^4(\eta)}\sum _{ij}
\left \langle \bar{h}_{ij}'(\eta ,\bm{x})
\bar{h}_{ij}'(\eta ,\bm{x})\right \rangle \, .
\end{equation}
where terms proportional to the Hubble constant have been neglected,
appropriate for $k/a\gg H$. Using the Fourier expansion, one obtains
\begin{eqnarray}
\label{eq:rhogw}
\rho _{\rm gw}(\eta )
&=&\frac{1}{4\kappa a^4(\eta)}\sum _{ij}
\int \frac{{\rm d}\bm{k}}{(2\pi )^{3/2}}
\int \frac{{\rm d}\bm{k}'}{(2\pi )^{3/2}}
{\rm e}^{-i\left(\bm{k}-\bm{k}'\right)\cdot \bm{x}}
kk' \Biggl\{
\left \langle A_{ij}(\bm{k})
A_{ij}^*(\bm{k}')\right \rangle \cos \left[k\left(\eta-\eta _{\rm f}
\right)\right]\cos \left[k'\left(\eta-\eta _{\rm f}
\right)\right] \nonumber \\ & &
-\left \langle A_{ij}(\bm{k})
B_{ij}^*(\bm{k}')\right \rangle \cos \left[k\left(\eta-\eta _{\rm f}
\right)\right]\sin \left[k'\left(\eta-\eta _{\rm f}
\right)\right]
-\left \langle B_{ij}(\bm{k})
A_{ij}^*(\bm{k}')\right \rangle \sin \left[k\left(\eta-\eta _{\rm f}
\right)\right]\cos \left[k'\left(\eta-\eta _{\rm f}
\right)\right]
\nonumber \\ & & 
+\left \langle B_{ij}(\bm{k})
B_{ij}^*(\bm{k}')\right \rangle \sin \left[k\left(\eta-\eta _{\rm f}
\right)\right]\sin \left[k'\left(\eta-\eta _{\rm f}
\right)\right]
\Biggr\}\, .
\label{rhogw2}
\end{eqnarray}
where we have also used $A_{ij}(\bm{k}) = A^*_{ij}(-\bm{k})$ and
$B_{ij}(\bm{k}) = B^*_{ij}(-\bm{k})$ imposed by $h_{ij}(\eta,\bm{x})$
being real. Expression Eq.~(\ref{rhogw2}) contains four terms. Let us
first discuss the first one. One needs to express the correlator of
two coefficients $A_{ij}$. Using the expressions established before,
see Eq.~(\ref{eq:A}), one is led to
\begin{eqnarray}
\left \langle A_{ij}(\bm{k})
A_{ij}^*(\bm{k}')\right \rangle =\frac{4\kappa ^2}{kk'}
\int _{\eta_{\rm ini}}^{\eta _{\rm f}}{\rm d}\tau 
\int _{\eta_{\rm ini}}^{\eta _{\rm f}}{\rm d}\tau '
\cos \left[k\left(\eta _{\rm f}-\tau
\right)\right]\cos \left[k'\left(\eta _{\rm f}-\tau'
\right)\right]a(\tau)a(\tau ')
\left \langle T_{ij}^{_{\rm TT}}(\tau, \bm{k})
T_{ij}^{_{\rm TT}*}(\tau', \bm{k}')\right \rangle\, ,
\end{eqnarray}
where one has
\begin{eqnarray}
\left \langle T_{ij}^{_{\rm TT}}(\tau, \bm{k})
T_{ij}^{_{\rm TT}*}(\tau', \bm{k}')\right \rangle
=\perp _{ij\ell m}\perp_{ijrs}
\left \langle T_{\ell m}(\tau, \bm{k})
T_{rs}^*(\tau', \bm{k}')\right \rangle \, ,
\end{eqnarray}
with the projectors defined in Eqs.~(\ref{eq:defperp})
and~(\ref{eq:defP}).

\par

To go further, one needs to specify what the stress-energy tensor
is. We take the stress-energy tensor of a pressure-less fluid since,
when the field rapidly oscillates at the bottom of its potential, the
pressure vanishes on average. Therefore, we have
\begin{equation}
\label{eq:EMtensor}
T_{ij}=\rho a^2 v_iv_j \, ,
\end{equation}
where $\bm{v}$ represents the velocity of the particles in the
pressure-less fluid. Note that Eq.~(\ref{eq:EMtensor}) is already
second order in the perturbation since $v_i$ is first order. The
velocity can be calculated from the continuity equation which reads
\begin{equation}
\label{eq:Euler}
\frac{\partial }{\partial t}\left(\frac{\delta \rho}{\rho}\right)
+\frac{1}{a}\nabla \cdot \bm{v}=0\, .
\end{equation}
Fourier transforming Eq.~(\ref{eq:Euler}), and noting that
$\delta\rho_{\bm{k}} /\rho\propto a(t)$ for growing density
fluctuations in which we are interested (as was discussed in the
previous section), we may relate the Fourier amplitudes of density
perturbation and velocity via
\begin{equation}
\bm{v}(\eta ,\bm{k})=-\frac{iHa}{k}
\delta_{\bm{k}}(\eta)\bm{e}_{\bm{k}}\, .
\end{equation}
Then, inserting the above expression for the velocity into the
definition of the stress-energy tensor Fourier component, one obtains
that
\begin{eqnarray}
\left \langle T_{\ell m}(\tau, \bm{k}) T_{rs}^*(\tau',
\bm{k}')\right \rangle &=& \frac{1}{(2\pi
  )^3}\rho(\tau)a^4(\tau)H^2(\tau) \rho(\tau')a^4(\tau')H^2(\tau')
\int {\rm d}\bm{q}\int {\rm d}\bm{p}
\frac{q_{\ell}(k_m-q_m)}{q^2(k-q)^2}
\frac{p_{r}(k_s'-p_s)}{p^2(k'-p)^2} \nonumber \\ & & \times
\left\langle \delta_{\bm{q}} (\tau)\delta_{\bm{k}-\bm{q}}
(\tau) \delta_{\bm{p}}^* (\tau')\delta_{\bm{k}'-\bm{p}}^*
(\tau') \right \rangle \, .
\end{eqnarray}
Then we use the fact that the density contrast grows proportional to
$a$ when the mode is inside the Hubble radius, while it remains
constant from the end of inflation until Hubble radius
crossing. Therefore, in the sub-horizon regime, one can write
$\delta_{\bm{q}} (\tau) = \delta_{\bm{q}} (\tau _{\rm end})
a(\tau)/a_{\rm hc}$, with $a_{\rm hc}$ the scale factor at horizon
crossing and $\delta_{\bm{q}} (\tau _{\rm end})$ the value of
$\delta_{\bm{q}}$ at the end of inflation. In order to match both
asymptotic behaviors in the super-horizon, namely $\delta_{\bm{q}}
(\tau)\rightarrow \delta_{\bm{q}} (\tau _{\rm end}) = (6/5)
\zeta_{\bm{q}}$ and in the sub-horizon regimes ,
$\delta_{\bm{q}}(\tau)\rightarrow
(2/5)\left[q/(aH)\right]^2\zeta_{\bm{q}}$, see Eq.~(\ref{eq:deltak}),
one defines $a_{\rm hc}$ as that at which $ q = \sqrt{3} a_{\rm
  hc}H_{\rm hc}$. Given that $ a H \propto a^{-1/2}$ in this
preheating stage, this indeed implies that $\left[q/(aH)\right]^2=3
(a/a_{\rm hc})$, hence the two asymptotes are matched at $a=a_{\rm
  hc}$.  Hence, one can write
\begin{equation}
\left \langle T_{\ell m}(\tau, \bm{k})
T_{rs}(\tau', \bm{k}')\right \rangle =
\frac{1}{(2\pi )^3}\frac{1}{(3\kappa )^2}
\int {\rm d}\bm{q}\int {\rm d}\bm{p}\,
q_{\ell}(k_m-q_m)p_{r}(k_s'-p_s)
\left\langle \delta_{\bm{q}} (\tau_{\rm end})
\delta_{\bm{k}-\bm{q}} (\tau_{\rm end}) 
\delta_{\bm{p}}^*(\tau_{\rm end})
\delta_{\bm{k}'-\bm{p}}^*(\tau_{\rm end})
\right \rangle \, ,
\end{equation}
The next step consists in evaluating the correlator. Using the Wick
theorem in order to express the four-point correlation function in
terms of three two-point correlations functions and discarding a
homogeneous piece we are not interested in, one arrives at
\begin{equation}
\left \langle T_{\ell m}(\tau, \bm{k})
T_{rs}(\tau', \bm{k}')\right \rangle =
\frac{2}{(2\pi )^3}\frac{1}{(3\kappa )^2}
\int {\rm d}\bm{q}\,
q_{\ell}(k_m-q_m)q_{r}(k_s-q_s)
\sigma ^2_{q}\sigma ^2_{\vert\bm{k}-\bm{q}\vert}
\delta(\bm{k}-\bm{k}')\, ,
\end{equation}
where we have defined $\left \langle \delta_{\bm{k}}(\tau_{\rm end})
  \delta^*_{\bm{k}'}(\tau_{\rm end})\right \rangle \equiv \sigma
^2_k\delta(\bm{k}-\bm{k}')$. Finally, we act with the projectors on
the last expression and one obtains
\begin{eqnarray}
\sum _{ij}\left \langle T_{ij}^{_{\rm TT}}(\tau, \bm{k})
T_{ij}^{_{\rm TT}*}(\tau', \bm{k}')\right \rangle &=&
\frac{1}{(2\pi )^3}\frac{1}{(3\kappa )^2}
\int {\rm d}\bm{q}\, q^4\sin^4 \alpha\, 
\sigma ^2_q\sigma ^2_{\vert\bm{k}-\bm{q}\vert}\delta(\bm{k}
-\bm{k}')\, ,
\end{eqnarray}
where $\alpha $ is the angle between the vectors $\bm{k}$ and
$\bm{q}$. Putting everything together, one can therefore evaluate the
correlator between the coefficients $A_{ij}$. We find
\begin{eqnarray}
\sum _{ij}\left \langle A_{ij}(\bm{k})
A_{ij}^*(\bm{k}')\right \rangle &=&\frac{4\kappa ^2}{kk'}
\int _{\eta_{\rm ini}}^{\eta _{\rm f}}{\rm d}\tau 
\int _{\eta_{\rm ini}}^{\eta _{\rm f}}{\rm d}\tau '
a(\tau )\cos \left[k\left(\eta _{\rm f}-\tau
\right)\right]a(\tau^{\prime})\cos \left[k'\left(\eta _{\rm f}-\tau'
\right)\right]
\frac{1}{(2\pi )^3}\frac{1}{(3\kappa )^2}
\nonumber \\ & & \times
\int {\rm d}\bm{q}\, q^4\sin^4 \alpha\, 
\sigma ^2_q\sigma ^2_{\vert\bm{k}-\bm{q}\vert}
\delta(\bm{k}-\bm{k}')\, .
\end{eqnarray}
Therefore, the first term in Eq.~(\ref{eq:rhogw}) is given by
\begin{eqnarray}
\rho _{\rm gw}^{\left \langle AA\right\rangle}(\eta )
&=&\frac{\kappa}{a^4(\eta)}
\int \frac{{\rm d}\bm{k}}{(2\pi )^3}\,
\cos ^2\left[k\left(\eta -\eta _{\rm f}\right)\right]
\int _{\eta_{\rm ini}}^{\eta _{\rm f}}{\rm d}\tau 
\int _{\eta_{\rm ini}}^{\eta _{\rm f}}{\rm d}\tau '
a(\tau )\cos \left[k\left(\eta _{\rm f}-\tau
\right)\right]a(\tau^{\prime})\cos \left[k\left(\eta _{\rm f}-\tau'
\right)\right]
\nonumber \\ & & \times
\frac{1}{(2\pi )^3}\frac{1}{(3\kappa )^2}
\int {\rm d}\bm{q}\, q^4\sin^4 \alpha\, 
\sigma ^2_q\sigma ^2_{\vert\bm{k}-\bm{q}\vert}\, .
\end{eqnarray}
Taking the time average of the previous expression over a period
replaces $\cos ^2\left[k\left(\eta -\eta _{\rm f}\right)\right]$ with
$1/2$, and cancels the cross-terms $\langle A B^*\rangle$, $\langle B
A^*\rangle$. The term $\rho _{\rm gw}^{\left \langle BB\right\rangle}$
is similar to $\rho _{\rm gw}^{\left \langle AA\right\rangle}$ except
that the term $\cos \left[k\left(\eta _{\rm f}-\tau \right)\right]\cos
\left[k'\left(\eta _{\rm f}-\tau' \right)\right]$ is replaced by $\sin
\left[k\left(\eta _{\rm f}-\tau \right)\right]\sin \left[k'\left(\eta
    _{\rm f}-\tau' \right)\right]$. When combined, these two terms
lead to a present day energy density
\begin{eqnarray}
\rho _{\rm gw}(\tau_0 )
&=&\frac{\kappa}{2(2\pi)^6a^4_0}
\int {\rm d}\bm{k}
\int _{\eta_{\rm ini}}^{\eta _{\rm f}}{\rm d}\tau 
\int _{\eta_{\rm ini}}^{\eta _{\rm f}}{\rm d}\tau '
\cos \left[k\left(\tau-\tau'\right)\right]a(\tau )a(\tau^{\prime})
\frac{1}{(3\kappa )^2}
\nonumber \\ & & \times
\int {\rm d}\bm{q}\, q^4\sin^4 \alpha\, 
\sigma ^2_q\sigma ^2_{\vert\bm{k}-\bm{q}\vert} 
=\frac{1}{(2\pi)^59\kappa a^4_0}
\int {\rm d}k\,k^2{\cal J}(k){\cal I}(k)\, ,
\end{eqnarray}
where the functions ${\cal J}$ and ${\cal I}$ are defined as follows
\begin{eqnarray}
{\cal J}(k)&\equiv& \int _{\eta_{\rm ini}}^{\eta _{\rm f}}{\rm d}\tau 
\int _{\eta_{\rm ini}}^{\eta _{\rm f}}{\rm d}\tau '\,
\cos \left[k\left(\tau-\tau'\right)\right]a(\tau)a(\tau')\, ,
\quad
{\cal I}(k)\equiv \int {\rm d}\bm{q}\, q^4\sin^4 \alpha\, 
\sigma ^2_q\sigma ^2_{\vert\bm{k}-\bm{q}\vert}\, .
\end{eqnarray} 
The first integral can be done exactly. The result reads
\begin{eqnarray}
{\cal J}(k)&\,=\,&\frac{1}{16H_{\rm end}^2}\left(\frac{k}{a_{\rm
    end}H_{\rm end}}\right)^{-6}\Bigl\{8+x_{\rm f}^4+x_{\rm
  ini}^4-\nonumber\\ 
& & \quad\quad 2\left[4+4x_{\rm f}
x_{\rm ini}-2x_{\rm ini}^2+x_{\rm f}^2\left(x_{\rm ini}^2-2\right)
\right]\cos\left(x_{\rm f}-x_{\rm ini}\right)
+4\left(x_{\rm ini}-x_{\rm f}\right)\left(2+x_{\rm f}x_{\rm ini}\right)
\sin\left(x_{\rm f}-x_{\rm ini}\right)\Bigr\}\, ,
\label{Jint}
\end{eqnarray}
with $x\equiv 3t_{\rm end}^{2/3}t^{1/3}k/a_{\rm end}=2(a/a_{\rm
  end})^{1/2}k/(a_{\rm end}H_{\rm end})$.  Note that $x>1$ for
sub-Hubble modes.  The second integral reads
\begin{equation}
{\cal I}(k)= 2\pi\int _{q_{\rm min}}^{q_{\rm max}}{\rm d}q
\int _{-1}^{+1}{\rm d}\mu\, \sigma ^2_q
\sigma ^2_{\vert\bm{k}-\bm{q}\vert}\left(1-\mu^2\right)^2q^6\, .
\end{equation}
Using $\delta_{\bm{q}}(\tau_{\rm end}) = (6/5)\zeta_{\bm{q}}$ and
Eq.~(\ref{Pzeta}), we have
\begin{equation}
\sigma ^2_q=\frac{36}{25}\frac{2\pi^2}{q^3}{\cal P}_{\zeta}(q)\, ,
\end{equation}
where the power spectrum of the curvature perturbation is calculated
at the end of inflation, see Fig.~\ref{fig:primspectre}. As a consequence,
one obtains
\begin{equation}
{\cal I}(k)=8\pi^5 \left(\frac{36}{25}\right)^2
\int _{q_{\rm min}}^{q_{\rm max}}{\rm d}q\,
\int _{-1}^{+1}{\rm d}\mu\, {\cal P}_{\zeta}(q)
{\cal P}_{\zeta}\left(\sqrt{q^2+k^2-2kq\mu}\right)
\frac{\left(1-\mu^2\right)^2q^3}{\left(q^2+k^2-2kq\mu\right)^{3/2}}\, .
\end{equation}
Using the above expressions, one finally obtains the present day
contribution of these GWs to the critical density, \ie ${\rm d}\Omega
_{\rm gw}/{\rm d}\ln k = \kappa/(3H_0^2)\, {\rm d}\rho _{\rm
  gw,0}/{\rm d}\ln k$
\begin{eqnarray}
\frac{{\rm d}\Omega _{\rm gw}}{{\rm d}\ln
  k}=\frac{12}{625}\left(\frac{a_{\rm
    end}}{a_0}\right)^4\left(\frac{H_{\rm
    end}}{H_0}\right)^2\left(\frac{a_{\rm rh}}{a_{\rm end}}\right)^2
\hat k\hat{\cal J}(\hat k)\hat {\cal I}(\hat k)\, ,
\label{omega1}
\end{eqnarray}
recalling the definitions introduced before: $\hat k \equiv k/(a_{\rm
  end}H_{\rm end})$, $\hat q \equiv q/(a_{\rm end}H_{\rm end})$ and
introducing the dimensionless quantities
\begin{eqnarray}
{\hat{\cal J}}(\hat k)& \,\equiv\,& 16 H_{\rm end}^2 x_{\rm f}^{-4}\hat
k^{6}{\cal J}(k)\,=\,H_{\rm end}^2\, \hat k^{2}\left(\frac{a_{\rm
      rh}}{a_{\rm end}}\right)^{-2}{\cal J}(k)\ ,
\label{calhatJ}\\
\hat{\cal I}(\hat k)&\,\equiv\, & \int _{\hat q_{\rm min}}^{\hat
  q_{\rm max}}{\rm d}\hat q\,
\int _{-1}^{+1}{\rm d}\mu\, {\cal P}_{\zeta}(q)
{\cal P}_{\zeta}\left(\sqrt{q^2+k^2-2kq\mu}\right)
\frac{\left(1-\mu^2\right)^2\hat q^3}{\left(\hat q^2+\hat k^2-2\hat
  k\hat q\mu\right)^{3/2}}\ ,
\label{calhatI}
\end{eqnarray}
\end{widetext}
where, in the argument of the power spectrum, $q$ and $k$ must be
considered as function of $\hat q$ and $\hat k$ according to the
simple relations introduced above.

\par

Expression Eq.~(\ref{omega1}) may be further simplified when assuming
a standard thermal history, \ie radiation domination between the
reheating epoch and the epoch of matter radiation-equality shortly
before recombination. In this case
\begin{eqnarray}
  \frac{{\rm d}\Omega _{\rm gw}}{{\rm d}\ln
    k} &\simeq & 2.8 \times 10^{-16}\Omega_{\gamma,0}
\biggl(\frac{T_{\rm rh}}{\rm 10^9 GeV}\biggr)^{-4/3}
\nonumber \\ & & \times 
\hat k\, \hat{\cal J}(\hat k)\, \hat {\cal I}_{11}(\hat k)\, ,
\end{eqnarray}
where $\Omega_{\gamma,0}\,\simeq\, 5\times10^{-5}$ is the present
density parameter in radiation and where we have defined ${\cal
  I}_{11}(\hat k)$ by ${\cal I}_{11}(\hat k)\equiv 10^{22}{\cal
  I}(\hat k)$ since ${\cal I}(\hat k)$ is quadratic in ${\cal
  P}_\zeta$. When reheating occurs long after the end of inflation,
\ie  for $x_{\rm f}\approx x_{\rm rh}\gg x_{\rm ini}\approx x_{\rm
  end}$ which implies $a_{\rm rh}\gg a_{\rm end}$, the expression in
the curly brackets of Eq.~(\ref{Jint}) approaches asymptotically
$x_{\rm rh}^4$. In this case $\hat{\cal J}$ approaches unity.

\par

Finally, it may also be interesting to have an expression giving the
associated present day GW frequency.  One finds
\begin{equation}
  f_0 \simeq 5\times 10^5{\rm Hz}\,\hat{k}\, 
  \biggl(\frac{T_{\rm rh}}{\rm 10^9 GeV}\biggr)^{1/3}\, .
\end{equation}

\par

\begin{figure*}
  \centering
\begin{tabular}{cc}
  \includegraphics[width=0.5\textwidth,clip=true]{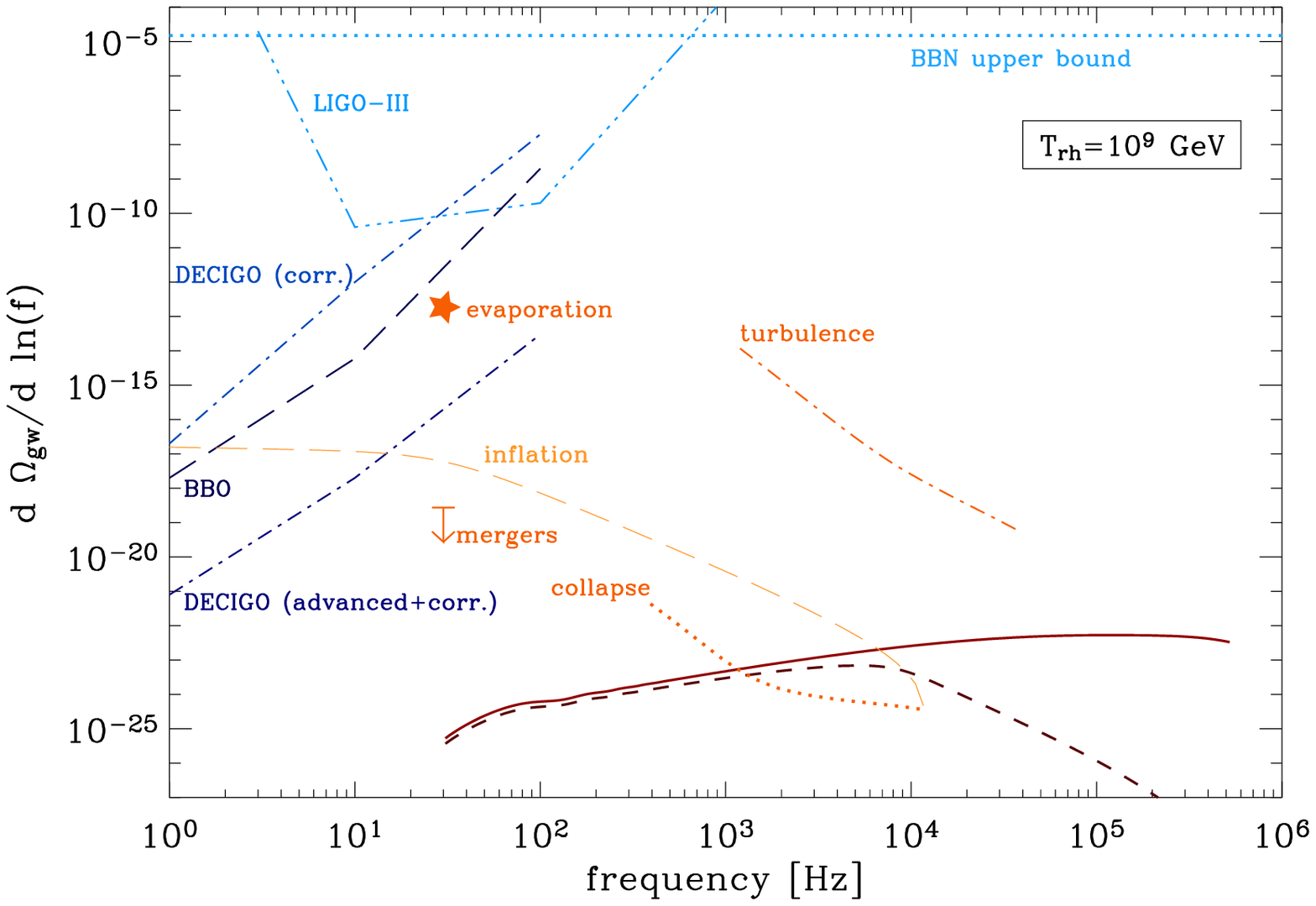}
  \includegraphics[width=0.5\textwidth,clip=true]{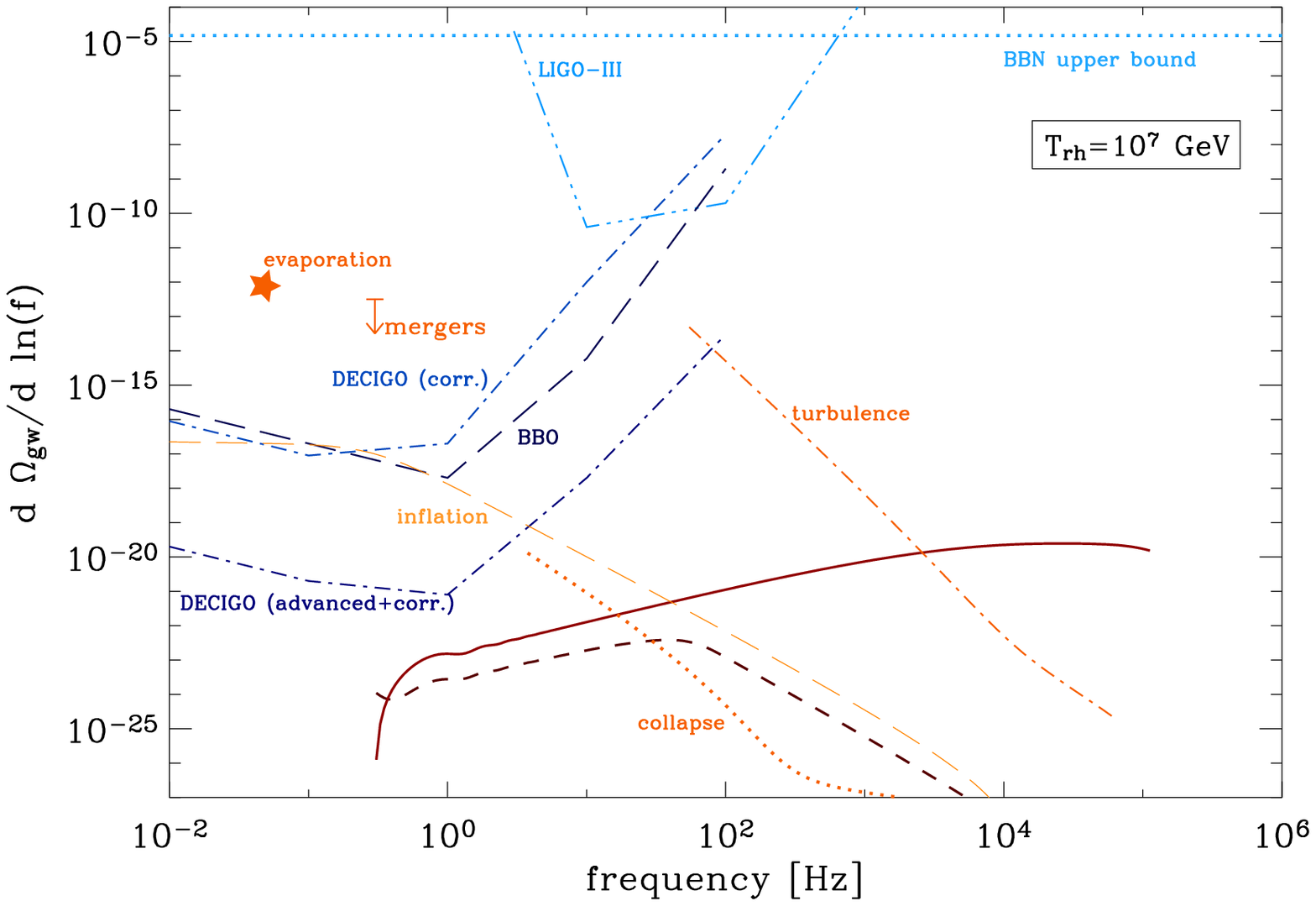}
\end{tabular}
  \caption[...]{Energy density in gravitational waves (in unit of the
    present day critical density) as a function of frequency. Left
    panel: $T_{\rm rh}=10^9\,$GeV; right panel: $T_{\rm
      rh}=10^7\,$GeV. The thick solid line shows the result of the linear
    calculation of the GW signal. The thick short dashed line shows the same
    calculation, but when stopped at the onset of the non-linear stage (see
    text). Various non-linear signal (discussed in
    Section~\ref{sec:nonlin}) are shown: the dotted line (``collapse'')
    shows the signal associated with the collapse of halos that have gone
    non-linear by reheating; the upper limit ``merger'' presents an
    estimate of the possible amplification of the collapse signal due
    to tidal interactions; the dashed-dotted line (``turbulence'') shows the
    signal expected from the turbulence produced by the evaporation of
    non-linear structures at reheating; and the star symbol
    (``evaporation'') shows the signal produced by the evaporation of the
    collapsed halos at reheating. The long dashed line annotated
    ``inflation'' shows the signal expected from the direct generation
    of GWs in $m^2\phi^2$ inflation. On the left hand side, the
    various lines indicate the projected sensitivities of next
    generation gravitational wave detectors. The horizontal dotted
    line gives the limit imposed by big-bang
    nucleosynthesis.} \label{fig:omegagw}
\end{figure*}

In Fig.~\ref{fig:omegagw}, we show the result of a numerical
integration of the above formula for two different reheating
temperatures $T_{\rm rh}=10^9\,$GeV and $T_{\rm
  rh}=10^7\,$GeV, respectively, along with projected
sensitivities of next generation gravitational wave experiments,
namely advanced LIGO~\cite{Abbott:2006zx,Smith:2009bx},
BBO~\cite{Corbin:2005ny} and DECIGO~\cite{Kawamura:2006up}. The
prediction for the above gravitational wave signal is shown by
the thick solid line. At this stage, it is important to stress that
the above calculation is carried out in the linear approximation and
cannot describe the non-linear aspects. In particular, there exists a
range of modes $[k_{\rm nl},k_{\rm max}]$ for which growth is
sufficiently effective that these modes become non-linear by the time
reheating starts. This range of wave numbers is discussed in the next
Section. Here, we wish to stress that the linear growth
$\delta_{\bm{k}}\propto a$, as used in the above calculation, is not
adequate anymore in the non-linear regime. This has impact on the
above calculation in the range $[k_{\rm nl},k_{\rm max}]$. In order to
show how the calculation could be affected, we have plotted in thick
dashed line in Fig.~\ref{fig:omegagw} the result obtained when one
shuts off the emission of gravitational waves when mode $k$ or
$\vert\bm{k}-\bm{q}\vert$ become non-linear. This curve, which sets a
strict lower limit to the gravitational wave signal departs from the
above linear calculation in the range $[k_{\rm nl},k_{\rm max}]$.

\par

In Fig.~\ref{fig:omegagw}, we also plot several estimates of
gravitational wave production in the non-linear regime, to be
discussed in Section~\ref{sec:nonlin}. We also plot the signal
expected from the direct production of gravitational waves in
$m^2\phi^2$ inflation (represented as the long dashed line annotated
``inflation''). We have calculated this signal following the study of
~\cite{Nakayama:2008wy}.  The change of spectral evolution (from
nearly scale invariant at small frequencies to $\propto k^{-2}$)
occurs at a frequency $k_{\rm min}/(2\pi)$, with $k_{\rm min}= a_{\rm
  rh}H_{\rm rh}$. Indeed, modes with $k>k_{\rm
  min}$ have re-entered the Hubble radius during inflaton domination
in the pre-reheating epoch and suffered redshifting while those modes
with $k<k_{\rm min}$ have re-entered the horizon in the radiation
dominated era (up to the modes with very small $k$ that re-enter in
the matter dominated era after matter-radiation equality, of course).

\par

It is clear that the magnitude of the linear signal is weak for the
reheating temperatures considered here. It is important to recall,
however, that such reheating temperatures lie in the upper range of
constraints set by the impact of gravitinos on big-bang
nucleosynthesis and that in the absence of a definite underlying
particle physics model in the inflaton sector, there is no particular
scale for $T_{\rm rh}$. The reheating temperature could be much lower,
in which case the strength of the signal would increase (roughly as
$T_{\rm rh}^{-4/3}$, see above) and the frequency would decrease (as
$T_{\rm rh}^{1/3}$), coming closer to the range of maximum sensitivity
of future instruments.

\section{Generation of Gravitational Waves in the 
Non-Linear Regime}
\label{sec:nonlin}

\subsection{Collapse of halos}
\label{subsec:collapse}

After perturbations have grown in the linear regime to $\delta\rho_{\bm k}
/\rho \sim 1$ they separate from the Hubble flow and collapse to form
halos. In order to calculate the gravitational wave signal coming from
this phase, we assume that the halos collapse and virialize
instantaneously when the fluctuation on the scale considered exceeds
the threshold $\delta_{\rm c}\simeq 1.7$. With this approximation, a
halo of mass $M$ emits gravitational waves at a well-defined time and
at a single pulsation $2\pi f$, which is estimated as the reciprocal
of the dynamical timescale of collapse $t_{\rm coll}$.

\par

The energy density emitted during the collapse of perturbations
corresponding to mass between $M$ and $M+{\rm d}M$ is
\begin{equation} 
{\rm d}\rho _{\rm gw}^{\rm coll}\,\simeq\,{\rm d}
n_{\rm h}^{\rm coll}{\cal L}_{\rm gw}^{\rm coll}t_{\rm
  coll}\left(\frac{a_{\rm coll}}{a_0}\right)^4 \, .\label{eq:drhocoll}
\end{equation}
In this expression, ${\cal L}_{\rm gw}^{\rm coll}$ is the luminosity
emitted by one halo of mass $M$ at collapse, ${\rm d}n_{\rm h}^{\rm
  coll}$ represents the number density of halos of mass between $M$
and $M+{\rm d}M$ at the time of collapse, and $t_{\rm
  coll}\,\simeq\,(\kappa\rho_{\rm h}/3)^{-1/2}$ is the collapse
timescale (with $\rho_{\rm h}$ the density of the halo at the time of
collapse, see below). The factor $\left(a_{\rm coll}/a_{\rm
    now}\right)^4$ accounts for the redshifting of the signal through
cosmic expansion. The value of $a_{\rm coll}$ as a function of
wave number will be determined further below. A halo of mass $M$
corresponds to a structure characterized by a comoving wave-number $k$
such that (assuming spherical symmetry)
\begin{equation}
  M=\frac{4\pi}{3}(2\pi)^3\frac{\rho _{\rm end}}{H_{\rm end}^3}
  \hat k^{-3}\, .\label{eq:Mofk}
\end{equation}
Therefore, the above equation can be rewritten as
\begin{equation}
  \frac{{\rm d}\Omega _{\rm gw}^{\rm coll}}{{\rm d}\ln f}
=\frac{\kappa}{3H_0^2}\frac{{\rm d}n_{\rm h}^{\rm coll}}{{\rm d}\ln M}
\frac{{\rm d}\ln M}{{\rm d}\ln f}{\cal L}_{\rm
    gw}^{\rm coll}t_{\rm coll}\left(\frac{a_{\rm coll}}{a_0}\right)^4\, ,
\end{equation}
with $f$ the comoving frequency of emission of gravitational waves,
$f\,\simeq\,(2\pi t_{\rm coll})^{-1}$, to be evaluated further below.

\par

The luminosity ${\cal L}_{\rm gw}^{\rm coll}$ can be evaluated as
follows~\cite{1984AmJPh..52..412S}. The typical amplitude of a
gravitational wave at point $\mathbf{x}$ (the origin of the
coordinates being chosen at the location of the emitting object) is
given by the quadrupole formula which reads
\begin{equation}
h\simeq \frac{G}{2}\left(\ddot{I}_{ij}-\frac{1}{3}
\ddot{I}_{kk}\delta _{ij}\right)
\frac{{n_i}{n_j}}{\vert \mathbf{x}\vert}\, ,
\end{equation}
where $n_i\equiv x_i/\vert \mathbf{x}\vert$. The quantity in brackets
is the trace free part of the quadrupole tensor $I_{ij}$. Since
spherical motions do not generate gravitational waves, we need to
assume that the halos are not spherically symmetric. A rough estimate
of the quantity $\ddot{I}_{ij}$ is
\begin{equation}
\ddot{I}_{ij}\simeq 2\int_{h} \rho_{\rm h}v_iv_j\simeq Mv^2\simeq 
\frac{2GM^2}{R}\, ,
\end{equation}
where $R$ is the typical size of the halo (physical radius) at
virialization and the virialized velocity $v_{\rm vir}\simeq
\sqrt{2GM/R}$. The luminosity of the source is thus given by
\begin{equation}
{\cal L}_{\rm gw}^{\rm coll}\simeq \frac{\vert \mathbf{x}\vert ^2}{G}
\dot{h}^2\simeq \frac{G^4M^5}{R^5}\, ,
\end{equation}
where we have written $\dot{h}\simeq h/T$ with $T\simeq R/v_{\rm
  vir}$. The radius $R$ is related to the wavenumber of the
fluctuation by $R\equiv 2\pi a_{\rm coll}/k$.

\par

Formally, the number density of collapsed halos at a given time can be
derived from a Press-Schechter mass function. With the above
approximation at that all halos of mass $M$ collapse instantaneously
when the rms mass fluctuation on that scale exceeds $\delta_{\rm c}$,
one derives the number density of halos of mass between $M$ and
$M+{\rm d}M$ at collapse as
\begin{equation} 
{\rm d}n_{\rm h}^{\rm
    coll}\,\simeq\,\frac{\bar\rho(a_{\rm coll})}{M}\,{\rm d}\,{\rm ln}\,M
  \,\simeq\,\frac{\bar\rho(a_{\rm end})}{M}\left(\frac{a_{\rm
        end}}{a_{\rm coll}}\right)^{3}\,{\rm d}\,{\rm ln}\,M\ ,
\end{equation}
with $\bar\rho(a_{\rm coll})$ the mean density of the Universe at the
time of collapse. This relation is nothing but the differential
version of the equation $n_{\rm h}/V=\bar{\rho}/M$ ($V$ is the volume)
which expresses the fact that, according to Press-Schechter, at a
given time, most of the matter is contained in the halos that
collapse at that time.

\par

The rms mass fluctuation $\sigma_M$ is related to $\sigma_k$, the
root-mean-square of the Fourier amplitude of density perturbations via
\begin{equation}
  \sigma_M^2 \simeq
  \int_0^{k} \frac{{\rm d^3} k'}{(2\pi)^3}
  \,\sigma^2_{k'}\, ,
\label{sigmaM}
\end{equation}
with $k$ the comoving wave number of the perturbation enclosing a mass
$M$ defined in Eq.~(\ref{eq:Mofk}). In the last equation an
approximation of simple k-space filtering has been applied, \ie only
modes $k'\leq k$ contribute to the average.  Using Eq.~(\ref{sigmaM})
one finds
\begin{equation}
  \sigma_M\approx\frac{1}{5}{\cal P}_{\zeta}(k)^{1/2}\frac{a}{a_{\rm
      end}}\hat k^2\, .
\label{eq.120}    
\end{equation}
Collapse of mass scale $M$ occurs when $\sigma_M\,\simeq\,\delta_{\rm
  c}$, hence at a collapse scale factor $a_{\rm coll}$ defined by
\begin{equation}
  \frac{a_{\rm coll}}{a_{\rm end}}\,=\, 5\delta_{\rm c} {\cal
    P}_\zeta(k)^{-1/2} \hat k^{-2}\ .\label{eq:acoll}
\end{equation}
At collapse, the typical over-density $\delta_{\rm coll}\,\sim\,150$
(in the spherical collapse model).  The present day frequency is
related to the comoving wave number of the perturbation as
\begin{eqnarray}
  f_0&\,=\,&\frac{1}{2\pi}\frac{a_{\rm coll}}{a_0}H_{\rm coll}
\left(1 + \delta_{\rm
      coll}\right)^{1/2}\ ,\nonumber\\
&\,\simeq\,& 3.8\times 10^3\,{\rm Hz}\, \hat k\,
\left(\frac{T_{\rm rh}}{10^9\,{\rm GeV}}\right)^{1/3}
\left(\frac{{\cal P}_\zeta}{10^{-11}}\right)^{1/4}
\ .
\end{eqnarray}
Combining all the above, one ends up with
\begin{eqnarray}
  \frac{{\rm d}\Omega_{\rm gw}^{\rm coll}}{{\rm d}\ln f}
&\,\sim\,& 6\times 10^3
  \Omega_{\gamma,0}\frac{a_{\rm end}}
{a_{\rm rh}}{\cal P}_{\zeta}^{5/4}\,\nonumber\\
  &\,\sim\,& 2\times10^{-23}\hat k^{-2}\left(\frac{T_{\rm rh}}{10^9\,{\rm
      GeV}}\right)^{4/3}\left(\frac{{\cal P}_\zeta}
{10^{-11}}\right)^{5/4}\ .\label{eq:116}
\end{eqnarray}
Maximum emission is provided by the largest scales that go non-linear
at reheating, i.e those with wave number $k_{\rm nl}$ such that $a_{\rm
  coll}(k_{\rm nl}) = a_{\rm rh}$. According to Eq.~(\ref{eq:acoll}),
this corresponds to
\begin{equation}
\hat k_{\rm nl}\,\simeq\, 0.1\left(\frac{T_{\rm rh}}{10^9\,{\rm
      GeV}}\right)^{2/3}\left(\frac{{\cal P}_\zeta}{10^{-11}}\right)^{-1/4}\ ,
\end{equation}
where we approximated ${\cal P}_{\zeta}(k_{\rm nl})=10^{-11}$ in
accord with Fig.~\ref{fig:primspectre}. The frequency at maximum
emission is thus
\begin{equation}
f_0^{\rm coll}\simeq 3.7 \times 10^2\, {\rm Hz}\, 
\biggl(\frac{T_{\rm rh}}{10^9\,{\rm GeV}}\biggr)\, .
\label{eq:freq}
\end{equation}
This falls in an interesting range considering the LIGO or BBO
gravitational wave detectors.  Nevertheless, the signal is fairly
weak: at the frequency of peak emission, one finds ${\rm d}\Omega_{\rm
  gw}^{\rm coll}/{\rm d}\,{\rm ln}\,f\,\sim 10^{-20}$ independent of
reheating temperature, a signal which would be only very hard to
detect.  The signal as a function of frequency is represented in
Fig.~\ref{fig:omegagw} in dotted line, annotated ``collapse''.

\subsection{After initial collapse of halos}
\label{subsec:aftercoll}

We have so far assumed that efficient GW emission in the non-linear
regime happens only during the collapse phase of halos, up to the
point where virialization is completed. It is not clear if
post-virialization rotation/vibration of halos, potentially enhanced
by occasional tidal forces acting on the halos due to passing by of
other halos, may lead to further efficient emission of GWs. The
question of how much more GWs are emitted after halo formation may
only be addressed by complete numerical simulations and is beyond the
scope of this paper. However, we may attempt to estimate it,
introducing an efficiency parameter $\varepsilon < 1$, for emission of
GWs after virialization. We may formally relate the generated GWs at
times between the end of collapse and reheating $\rho_{\rm gw}^{>\,
  \rm coll}$ to $\rho_{\rm gw}^{\rm coll}$ in Eq.~(\ref{eq:drhocoll})
by making the following replacements in Eq.~(\ref{eq:drhocoll}):
$\rho_{\rm gw}^{\rm coll,0} \to \rho_{\rm gw}^{>\, \rm coll,0}$, ${\rm
  d}n_h(M) \to {\rm d}n_h^{>\,\rm coll}(M) = {\rm d}n_h(M) (a_{\rm
  coll}/a)^{3}$ to take into account the fact that the signal is
emitted after the collapse, ${\cal L}_{\rm gw}^{\rm coll} \to {\cal
  L}_{\rm gw}^{\rm >\, coll} = \varepsilon {\cal L}_{\rm gw}^{\rm
  coll}$ as discussed above, and, finally, $t_{\rm coll} \to t_{>\,\rm
  coll} = t_{\rm coll}(a/a_{\rm coll})^{3/2}$, $(a_{\rm coll}/a_0)^{4}
\to (a/a_0)^{4}$, again to take into account that the emission time is
now different. Inserting this into Eq.~(\ref{eq:drhocoll}) yields
\begin{equation}
{\rm d}\rho_{\rm gw}^{>\, \rm coll,0} \approx 
{\rm d}\rho_{\rm gw}^{\rm coll,0}
\biggl(\frac{a}{a_{\rm coll}}\biggr)^{5/2}\varepsilon\, ,
\end{equation}
implying a potentially large enhancement of the signal when $a_{\rm
  rh}\gg {a_{\rm coll}}$, provided $\varepsilon$ is not too small.
Using this in conjunction with Eq.~(\ref{eq:116}) one finds that for
non-negligible $\varepsilon$ the signal may be dominated by the emission
of GWs at $T_{\rm rh}$ from the first structures which had formed. One
would then expect a signal ${\rm d}\Omega_{\rm gw}/{\rm d}\,{\rm
  ln}\,f\sim 2\times10^{-18} \varepsilon\,(T_{\rm rh}/10^9\,{\rm
  GeV})^{-2}({\cal P}_\zeta/10^{-11})^{5/2}$ with a rather strong
dependence on the reheating temperature, which would make it
potentially detectable (provided $\varepsilon$ is not too small, of
course): for $T_{\rm rh}\sim 10^6\,$GeV, the signal becomes of order
$10^{-12}$ in the $10^{-2}\,$Hz range. Assuming indeed that the
typical pulsation corresponds to the Hubble scale at the time of
emission, one finds a frequency $f_0^{>{\rm coll}}\sim 30\,{\rm
  Hz}\,(T_{\rm rh}/10^9\,{\rm GeV})$ for the peak emission at $T_{\rm
  rh}$. This signal is indicated by the ``merger'' upper limit in
Fig.~\ref{fig:omegagw}, given the uncertainty on the value of
$\varepsilon$.

\subsection{Evaporation of halos}
\label{subsec:evaporation}

Around the epoch of reheating, the self-gravitating halos evaporate
due to emission of radiation by the decaying inflaton. Due to the
smallness of the inflaton-radiation coupling, the radiation does not
scatter on the inflaton particles contained in the halos, but escapes
freely without communicating its pressure to the inflaton halo. In
this picture, each halo behaves as a source of radiation with a
typical emission timescale $\tau_\phi = \Gamma_\phi^{-1}$
corresponding to the inflaton lifetime. The mean free path of
radiation against scattering with itself is
$\lambda\,\sim\,1/(\alpha^2 T)$, with $\alpha$ a typical coupling
constant and $T$ the temperature of the radiation, therefore
$\lambda\,\ll\,\tau_{\phi}$ and the radiation can be approximated as
an instantaneously thermalized ultra-relativistic fluid. The radiation
emitted by the evaporating halo expands and accelerates under its own
internal pressure until it interacts with the winds emitted by
surrounding halos. We will address this phase in the following
subsection and for the time being, we estimate the amount of
gravitational waves produced during the evaporation of the halos,
taken individually.

\par

Since the radiation escapes on a timescale $t_{\rm esc}=R/c$, with $R$
the radius of the virialized halo, and since $R\,\ll\,H^{-1}$, the
hierarchy $t_{\rm esc}\,\ll\,\tau_{\phi}$ remains satisfied all
throughout evaporation. This implies that the energy of radiation
contained in the halo at any time, $M t_{\rm esc}/\tau_{\phi}$ is
always smaller than $M$, so that its gravitational influence can be
neglected.

\par

As the halo evaporates, its virial radius increases in inverse
proportion to $M$. In order to see this, one first notices that the
dynamical timescale of evolution of the halo, $t_{\rm
  dyn}=(\kappa\rho_{\rm h}/3)^{-1/2}$ remains much smaller than the
Hubble time $(\kappa\rho/3)^{-1/2}$ during the evaporation process,
since $\rho_{\rm h}\,\gg\,\rho$. Therefore the structure adjusts
itself on a timescale which is much smaller than the evaporation
timescale $\tau_{\phi}\,=\,H_{\rm rh}^{-1}$ ($H_{\rm rh}^{-1}$ the
Hubble time at reheating). In other words, the halo undergoes a series
of quasi-static equilibria as the mass decreases, the virial relation
$Mv^2 \,\simeq\, 2GM^2/R$ being verified at each step. Through
adiabatic expansion, the rms velocity decreases as $1/R$, hence one
can check that ${\rm d}M/M=-{\rm d}R/R$.

\par

During evaporation, the luminosity can be expressed
as~\cite{1984AmJPh..52..412S}
\begin{equation}
{\cal L}_{\rm gw}\,\simeq\, \bar{\varepsilon}G
\left(\frac{{\rm d}M}{{\rm
    d}t}\right)^2\ .
\label{eq:Idot}    
\end{equation}
Here $\bar{\varepsilon}<1$ is an efficiency factor which represents a
measure of the asphericity of the radiation wind, analogous to the
factor introduced in Section~\ref{subsec:aftercoll}. 
Using the previous arguments, the amount of energy density emitted in
gravitational waves on an evaporation timescale $\tau_\phi$
approximately reads:
\begin{eqnarray}
{\rm d}\rho_{\rm gw}&\,\simeq\,& \bar{\varepsilon}\, 
\frac{G}{4}\left(\frac{{\rm
    d}M}{{\rm d}t}\right)^2\tau_\phi\, 
{\rm d}n_{\rm h\vert rh}(M)\,\nonumber\\
    &\,\simeq\,& \bar{\varepsilon}\frac{G}{2} M^2 H_{\rm rh} 
{\rm d}n_{\rm h\vert rh}(M)
\end{eqnarray}
where ${\rm d}n_{\rm h\vert rh}$ is the number density of halos of
mass comprised between $M$ and $M+{\rm d}M$ at the time of reheating
and $H_{\rm rh}\,\simeq\,1/(2\tau_\phi)$ denotes the Hubble scale at
reheating. It is important to note that all halos evaporate on the
same timescale $\tau_\phi$, therefore they all contribute to the
gravitational wave signal at the same comoving pulsation $\sim a_{\rm
  rh} \tau_\phi^{-1}$. Due to the quadratic dependence on mass scale
$M$ the signal is dominated by the largest structures
evaporating. This is given by the typical mass scale which collapses
shortly before reheating. It may be obtained by employing
Eq.~(\ref{eq:acoll}) with $a_{\rm coll}\approx a_{\rm rh}$ yielding
the final simple result
\begin{equation}
  \Omega_{\rm gw}\sim \bar{\varepsilon}\,
\Omega_{\gamma,0}\, P_{\zeta}(k_{\rm nl})^{3/4}\, ,
\label{eq:Omega2}
\end{equation}
independent of reheating temperature. The typical frequency of
emission corresponds to the (redshifted) Hubble scale of reheating
(divided by $2\pi$) and is given as before by $f_0^{>{\rm coll}}\sim
30\,{\rm Hz}\,(T_{\rm rh}/10^9\,{\rm GeV})$, in an interesting range
from the point of view of detection.

\par

The above signal may be quite substantial, in particular, for
$P_{\zeta}\sim 10^{-11}$ as implied by Fig.~\ref{fig:primspectre} one
obtains $\Omega_{\rm gw}\sim 4\times 10^{-13}\bar{\varepsilon}$. Of
course, the signal may be smaller than our estimate in case velocities
are below the speed of light and/or $\bar{\varepsilon}\ll 1$.

\subsection{Dissipation after reheating}
\label{subsec:dissipation}

As explained above, once the radiation is emitted, it will shock
against the radiation generated by surrounding halos. All in all, this
injects a large amount of energy on all scales $\lesssim L_{\rm
  s}\,\equiv\,2\pi a_{\rm rh}/k_{\rm nl}$, where $k_{\rm nl}$ denotes
as before the smallest comoving wave number of perturbations that go
non-linear by reheating. On general grounds, one expects that this
injection of energy stirs a turbulent cascade on scales $\lesssim
L_{\rm s}$, which decays through damping at small scales. This cascade
is characterized by a power spectrum of kinetic energy in modes
(eddies) of various wave numbers, which can emit gravitational
radiation in the very same way that growing perturbations do during
the linear phase. This has been discussed in detail by Kosowsky {\it
  et. al}~\cite{Kosowsky:2001xp} and we simply adapt their calculation
to the present framework.  We may assume here that a fraction of order
unity of the total energy density at reheating is injected in the
cascade, \ie kinetic energy of the radiation fluid is of order the
rest energy. The stirring scale $L_{\rm s}$ can be evaluated by using
Eq.~(\ref{eq:acoll}), with $a_{\rm coll}=a_{\rm rh}$, which yields
\begin{equation}
L_{\rm s}\,\simeq\,2\,{\cal P}_\zeta^{1/4} H_{\rm rh}^{-1}\ .
\end{equation} 
The stirring scale is much smaller than the timescale $\tau_{\phi}\sim
H_{\rm rh}^{-1}$ over which energy is injected into the plasma, so
that one may consider the turbulence to be stationary over this
timescale $\tau_{\phi}$. Following~\cite{Kosowsky:2001xp}, one then
finds that the gravitational wave signal generated is:
\begin{eqnarray}
  \frac{{\rm d}\Omega_{\rm gw}}{{\rm d}\,{\rm ln}\,f}
  &\,\sim\,&2\times 10^{-7}\,\left(\frac{\tau_{\phi}}{H_{\rm
        rh}^{-1}}\right)^{-1}
  \left(\frac{L_{\rm s}}{H_{\rm
        rh}^{-1}}\right)^3 \left(\frac{f}{f_{\rm s}}\right)^{-7/2}\
  ,\nonumber\\
  &\,\sim\,&2 \times 10^{-14} \left(\frac{{\cal
        P}_{\zeta}}{10^{-11}}\right)^{3/4}
\left(\frac{f}{f_{\rm s}}\right)^{-7/2}\ ,
\end{eqnarray}
independent of reheating temperature. The typical frequency is defined
as
\begin{eqnarray}
  f_{\rm s}&\,=\,&\frac{1}{3\pi}\frac{a_{\rm rh}}{a_0}
\left(\tau_{\phi}L_{\rm s}^2\right)^{-1/3}\,\nonumber\\
  &\,=\,&  10^3\,{\rm Hz}\left(\frac{T_{\rm rh}}{10^9\,{\rm
        GeV}}\right)\left(\frac{{\cal P}_{\zeta}}{10^{-11}}\right)^{-1/6}\ .
\end{eqnarray}
Although not as strong as that resulting from the evaporation of
halos, this signal remains substantial and the typical frequency can
easily shift down to the interesting range for detection if the
reheating temperature $T_{\rm rh}\,\lesssim\,10^7\,$GeV. This signal
is indicated in Fig.~\ref{fig:omegagw} by the dashed-dotted line
annotated ``turbulence''.

\section{Conclusion}
\label{sec:conclusion}

In this paper we have calculated the gravitational wave signal
associated to the growth of sub-horizon perturbations between the end
of cosmic inflation and the beginning of a radiation dominated early
Universe in $m^2\phi^2$ chaotic inflation. The growth of metric
fluctuations formally arises from a preheating like instability and
can be interpreted, in the range of wave numbers of interest, as the
gravitational instability of a pressure-less fluid in a matter
dominated era. The over-density $\delta\rho_{\bm k} /\rho$ of these
perturbations grows linearly with scale factor and reach non-linearity
to re-collapse and form inflaton halos after only moderate expansion of
the Universe.

\par

Though these small-scale inflaton halos later evaporate when the
inflaton decays at reheating, they may nevertheless lead to the
emission of gravitational waves in an interesting frequency range
$\omega_0\sim 10^{-5}-10^8$Hz for current ground-based and planned
satellite gravitational wave detectors, such as Advanced LIGO, DECIGO
and BBO. We have therefore analyzed the gravitational wave emission in
detail during five distinct phases: (a) growth of the linear
perturbations, (b) first collapse of halos when reaching
non-linearity, (c) subsequent non-linear evolution of halos, (d)
evaporation of halos due to inflaton decay, and (e) cosmic turbulence
during the first epochs of radiation domination, respectively. We note
here that phase (a) has recently also been studied by
Ref.~\cite{Assadullahi:2009nf}.  An exact calculation of (a) indicates
that gravitational wave emission is weak, peaking at values of order
$\Omega_{\rm gw}\sim 10^{-22} T_{\rm rh}^{-4/3}$, if the magnitude of
curvature perturbations on small scales as predicted in chaotic
inflation, \ie $P_{\zeta}\sim 10^{-11}$, is assumed. The estimate of
gravitational wave emission during phase (b) is similarly small.
Promising gravitational wave signals of $\Omega_{\rm gw}\gtrsim
10^{-12}$ for $P_{\zeta}\sim 10^{-11}$, could however, result during
phases (c) (d) and (e), depending on the value of the reheating
temperature. Generally speaking, the lower the reheating temperature,
the longer the phase of growth of fluctuations and the larger the band
of wave numbers turning non-linear, and consequently, the larger the
signal of gravitational wave emission. The exact value of these
signals depends on the details of the structure formation process
including tidal disruption of substructures, rotation, asphericity,
etc. A reliable estimate may therefore only be given when full
numerical simulations are performed.

\bibliographystyle{h-physrev2}
\bibliography{biblio2}
\end{document}